\documentclass{article}
\usepackage[T1]{fontenc}
\usepackage{iclr2027_conference,times}

\usepackage{amsmath,amsfonts,bm}

\def\eqref#1{equation~\ref{#1}}

\def\1{\bm{1}}

\DeclareMathAlphabet{\mathsfit}{\encodingdefault}{\sfdefault}{m}{sl}
\SetMathAlphabet{\mathsfit}{bold}{\encodingdefault}{\sfdefault}{bx}{n}

\usepackage{hyperref}
\usepackage{url}
\usepackage{booktabs}
\usepackage{graphicx}
\usepackage{multirow}
\usepackage{amsmath}
\usepackage{algorithm}
\usepackage{algorithmic}
\usepackage{amssymb}
\usepackage{enumitem}
\usepackage[most]{tcolorbox}  
\usepackage[dvipsnames]{xcolor}

\definecolor{OurColor}{RGB}{235, 238, 242}
\usepackage[table]{xcolor}

\makeatletter
\@ifpackageloaded{amsmath}{}{\RequirePackage{amsmath}}
\@ifpackageloaded{amssymb}{}{\RequirePackage{amssymb}}
\@ifpackageloaded{xcolor}{}{\RequirePackage{xcolor}}
\@ifpackageloaded{tcolorbox}{}{\RequirePackage{tcolorbox}}
\@ifpackageloaded{tikz}{}{\RequirePackage{tikz}}
\makeatother

\tcbuselibrary{breakable,skins}
\usetikzlibrary{arrows.meta,positioning,fit,calc}

\definecolor{MASLoopCaseBlue}{RGB}{42,92,145}
\definecolor{MASLoopCaseBlueBg}{RGB}{241,247,253}
\definecolor{MASLoopCaseRed}{RGB}{166,55,55}
\definecolor{MASLoopCaseRedBg}{RGB}{253,243,243}
\definecolor{MASLoopCaseGreen}{RGB}{35,119,82}
\definecolor{MASLoopCaseGreenBg}{RGB}{239,249,244}
\definecolor{MASLoopCaseGold}{RGB}{161,112,21}
\definecolor{MASLoopCaseGoldBg}{RGB}{255,249,235}
\definecolor{MASLoopCaseGray}{RGB}{90,98,108}
\definecolor{MASLoopCaseGrayBg}{RGB}{247,248,250}

\newtcolorbox{masloopcaseoverview}[1]{
  enhanced, breakable,
  colback=MASLoopCaseBlueBg, colframe=MASLoopCaseBlue,
  boxrule=0.8pt, arc=1.5mm,
  left=1.5mm,right=1.5mm,top=1mm,bottom=1mm,
  fonttitle=\bfseries, title={#1}
}
\newtcolorbox{masloopfailurebox}[1]{
  enhanced, breakable,
  colback=MASLoopCaseGrayBg, colframe=MASLoopCaseGray,
  boxrule=0.8pt, arc=1.5mm,
  left=1.5mm,right=1.5mm,top=1mm,bottom=1mm,
  fonttitle=\bfseries, title={#1}
}
\newtcolorbox{masloopsuccessbox}[1]{
  enhanced, breakable,
  colback=MASLoopCaseGreenBg, colframe=MASLoopCaseGreen,
  boxrule=0.8pt, arc=1.5mm,
  left=1.5mm,right=1.5mm,top=1mm,bottom=1mm,
  fonttitle=\bfseries, title={#1}
}
\newtcolorbox{masloopevidencebox}[1]{
  enhanced, breakable,
  colback=MASLoopCaseGoldBg, colframe=MASLoopCaseGold,
  boxrule=0.8pt, arc=1.5mm,
  left=1.5mm,right=1.5mm,top=1mm,bottom=1mm,
  fonttitle=\bfseries, title={#1}
}
\newtcolorbox{maslooptracebox}[1]{
  enhanced, breakable,
  colback=MASLoopCaseGrayBg, colframe=MASLoopCaseGray,
  boxrule=0.6pt, arc=1mm,
  left=1.5mm,right=1.5mm,top=1mm,bottom=1mm,
  fonttitle=\bfseries, title={#1}
}

\tikzset{
  masloop-case-node/.style={draw, rounded corners, align=center,
    font=\scriptsize, text width=4.0cm, minimum height=9mm},
  masloop-case-flow-red/.style={-{Latex[length=1.8mm]}, thick, draw=MASLoopCaseRed},
  masloop-case-flow-green/.style={-{Latex[length=1.8mm]}, thick, draw=MASLoopCaseGreen}
}

\tikzset{
  masloop-case-node/.style={draw, rounded corners, align=center,
    font=\scriptsize, text width=4.0cm, minimum height=9mm},
  masloop-case-flow-red/.style={-{Latex[length=1.8mm]}, thick, draw=MASLoopCaseRed},
  masloop-case-flow-green/.style={-{Latex[length=1.8mm]}, thick, draw=MASLoopCaseGreen}
}

\newcommand{\refgain}[2]{#1\,{\scriptsize\textcolor{MASLoopCaseGreen}{(+#2)}}}
\newcommand{\revise}[1]{\textcolor{black}{#1}}

\usepackage{hyperref}      
\usepackage{fontawesome5}  
\usepackage{twemojis}      
\usepackage{cuted} 

\title{MAS-Loop: Beyond One-Shot Orchestration towards Self-Evolving Multi-Agent Systems}
\author{
\bf Hehai Lin\textsuperscript{$^{\spadesuit}$}, ~
Qi Yang\textsuperscript{$^{\diamondsuit}$}, ~
Chengwei Qin\textsuperscript{$^{\spadesuit}$} \thanks{Corresponding to \href{mailto:chengweiqin@hkust-gz.edu.cn}{chengweiqin@hkust-gz.edu.cn}}
\\
$^{\spadesuit}$The Hong Kong University of Science and Technology (Guangzhou),
$^{\diamondsuit}$Ant Group
}

\iclrfinalcopy 
\begin{document}
\maketitle

\vspace{-2em}
\begin{center}
  \href{https://linhh29.github.io/blog/MAS-Loop/index.html}{\twemoji{brain}\,Project page}
  \hspace{0.75em}
  \href{https://mas-loop-demo.hehailin.life/}{\textcolor{red}{\faYoutube}\,Online Demo}
  \hspace{0.75em}
  \href{https://github.com/linhh29/MAS_Loop}{\faGithub\,Code}
\end{center}

\begin{abstract}
Training-time multi-agent system (MAS) design seeks to learn reusable orchestration policies, yet prevailing approaches remain largely one-shot: an Orchestrator generates a system but only observes and learns from its terminal outcome. This paradigm discards execution trajectories that expose system-level failures and contain fine-grained information beyond the terminal outcome, leaving no direct basis for revising the source MAS, evaluating the updated MAS, or learning from the resulting contrast. To this end, we introduce \textbf{MAS-Loop}, an execution-grounded framework that closes this feedback loop by coupling an Orchestrator with a trajectory-conditioned Refiner. For each task, the Orchestrator constructs and executes a source MAS to obtain an outcome score and an intermediate trajectory. Conditioned on the trajectory, the Refiner localizes an actionable failure, formulates a trajectory-grounded diagnosis, and proposes a bounded patch to the agents, information flow, or execution configuration. The executable updated MAS is re-executed on the same task to measure its realized outcome under stochastic execution. This pair-MAS comparison provides complementary learning signals: the Orchestrator learns to generate strong systems with fewer recoverable failures, whereas the Refiner learns to expose and repair observed weaknesses, thereby driving policy-level co-evolution across training iterations. We evaluate MAS-Loop across interaction environments, knowledge-intensive multi-hop reasoning, and general reasoning, examining both the complete framework and the independent utility of these two policies. Experiments show that MAS-Loop achieves the strongest overall performance among the evaluated baselines, with consistent gains from refinement across all evaluated benchmarks. The co-evolution process also yields \textbf{MAS-JudgeCorpus}, a dataset of execution-grounded source-updated pairs with orchestration- and trajectory-level supervision, which can provide the community with support for MAS reward modeling and trajectory-aware investigation.

\end{abstract}

\section{Introduction}
\label{sec:introduction}

Training-time multi-agent system (MAS) design has emerged as a promising alternative to inference-time MAS design, where the workflow search must be repeated for different tasks and cannot accumulate reusable experience across tasks~\citep{zhang2025aflow, ke2025mas, hu2025automated, lin2026unified}. Training-time MAS trains an Orchestrator to construct a task-specific MAS by selecting models or tools, assigning agent roles, and organizing communication~\citep{li2024survey, ke2025survey, tran2025multi}. The system then executes and uses its outcomes to improve the Orchestrator through supervised learning~\citep{ye2025mas, yue2025masrouter}, reinforcement learning~\citep{yang2026learning, ke2026mas}, and graph-based optimization~\citep{zhuge2024language, zhang2024g}. In this way, experience acquired from previous executions can be encoded into a reusable orchestration policy and transferred to future tasks~\citep{lin2026skill}.


Although this makes MAS construction learnable, this \emph{one-shot orchestration} remains an incomplete foundation for effective MAS self-evolution~\citep{qi2026beyond}: orchestration and execution feedback remain disconnected. A system trajectory can expose missing information flow, redundant reasoning, or failures in verification and aggregation~\citep{shen2025understanding, cemri2026multi}. Yet this trajectory-level evidence is often omitted from the terminal correctness signal~\citep{jwalapuram2026illusion}. Therefore, without a feedback-driven refinement mechanism, the Orchestrator must discover better designs indirectly through independently sampled rollouts and their sparse outcome rewards~\citep{zhang2026reinforcement, su2025toolorchestra, lin2026mas, peng2026degrees}. Although this trial-and-error process may eventually favor stronger systems, it relies on vanilla independent rollouts and sparse terminal rewards rather than utilizing rich intermediate trajectories as feedback to directly diagnose failures, guide targeted repairs, and optimize from verified improvements.

\begin{figure*}[t]
    \centering\includegraphics[width=\linewidth]{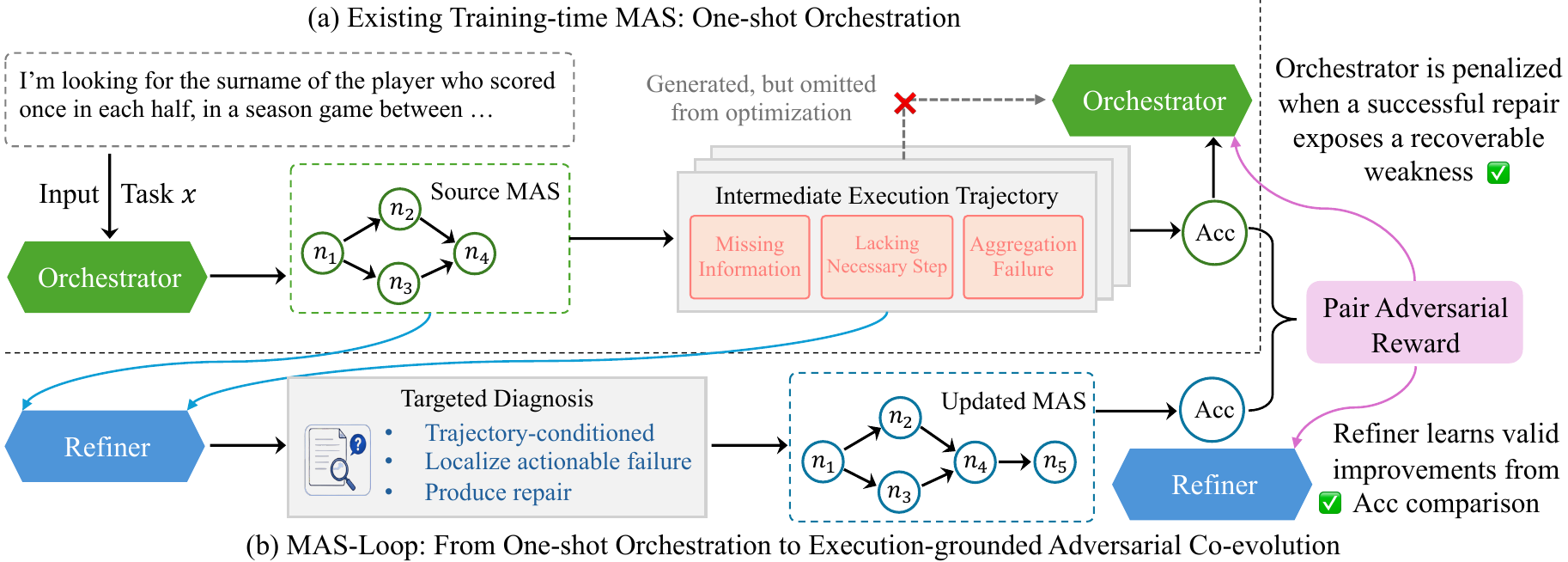}
    \caption{The Illustration of MAS-Loop. (a) Existing training-time MAS methods rely on one-shot orchestration and omit intermediate execution feedback from optimization. (b) MAS-Loop moves from one-shot orchestration to execution-grounded adversarial co-evolution by adding a trajectory-conditioned Refiner and incorporating coupled incentive learning signals.}
    \label{fig:background}
\end{figure*}

However, a complete self-evolution loop requires more than learning to generate the next system: it should connect execution evidence to a direct intervention on the current system and facilitate learning based on the consequences of that intervention~\citep{chen2026recursive, yang2026self}. To this end, we introduce \textbf{MAS-Loop}, an execution-grounded framework that augments the \emph{Orchestrator} with a complementary learned policy, the \emph{Refiner}. The Orchestrator first constructs a source MAS, whose execution produces a terminal score and an intermediate trajectory. \revise{The Refiner observes the source design, final prediction, and trajectory to localize actionable weaknesses, formulate a grounded diagnosis, and propose a bounded patch to agent roles, information flow, or execution settings.} This patch is compiled into an updated MAS and re-executed on the same task to \revise{measure the realized outcome change}. The source-updated score difference provides coupled learning signals: it encourages the Orchestrator to produce strong systems with fewer recoverable weaknesses and trains the Refiner to generate valid revisions that repair observed failures, enabling the two policies to co-evolve from this rich-learning closed-loop. \revise{This co-evolution occurs across policy-optimization iterations: each rollout contains one bounded refinement, while subsequent task rollouts use the newly updated Orchestrator and Refiner.}


During policy co-evolution, MAS-Loop naturally produces large numbers of execution-grounded source-updated pairs. To obtain reliable comparisons, we retain tasks with a fixed rollout budget, discard ties and pairs with insufficient absolute or relative score gaps, remove locally inconsistent examples, and canonicalize Refiner operations before operation-aware deduplication. The resulting \textbf{MAS-JudgeCorpus} contains source and updated MAS instances whose quality differences are established by same-task re-execution, providing pairwise preferences for training MAS reward models with \revise{orchestration}- and execution-level supervision. Then we leverage MAS-JudgeRM for test-time ranking and selection among candidate MAS, while the retained trajectory evidence also offers a basis for future trajectory-conditioned judging and other execution-aware evaluations.

We evaluate MAS-Loop across interaction environments, knowledge-intensive multi-hop reasoning, and general reasoning, including both in-task performance and out-of-task transfer. MAS-Loop achieves the strongest overall performance among the evaluated single-agent and automatic MAS baselines. The Orchestrator remains effective independently, while the Refiner consistently improves both its source systems and those produced by other MAS methods across in-task and out-of-task evaluations. For MAS-JudgeCorpus, we train MAS-JudgeRM on pairwise preferences and compare it with general-purpose reward models, orchestration-specific reward models, and prompted LLM judges for selecting among a fixed set of candidate MAS. Together, these experiments assess both MAS-Loop's effectiveness as a self-evolving MAS framework and MAS-JudgeCorpus's value for test-time MAS selection.
Our contributions are summarized as follows:
\begin{itemize}[leftmargin=*]
    \item We identify a central limitation of current training-time MAS: one-shot orchestration leaves execution feedback disconnected from the MAS design, forcing the Orchestrator to improve indirectly through sparse terminal rewards rather than direct rich trajectory-level evidence.
    \item We introduce \textbf{MAS-Loop}, an execution-grounded framework that couples an Orchestrator with a trajectory-conditioned Refiner. The paired re-execution \revise{evaluates the updated MAS under the same task protocol} and provides complementary learning signals that drive both policies to co-evolve \revise{across training iterations}.
    \item We construct \textbf{MAS-JudgeCorpus} from execution-grounded source--updated MAS pairs, applying fixed-rollout quality filtering, conservative consistency checks, and operation-aware deduplication to obtain reliable preferences for training MAS reward models.
\end{itemize}

\section{Related Work}
\label{sec:related_work}

\subsection{Automatic Multi-Agent System Design}

Inference-time MAS design typically uses a powerful frontier LLM together with iterative search, reflection, or optimization to construct an agent workflow for each task~\citep{ruan2026aorchestra, ma2026self, zhou2025reso, zhou2026multi}. This approach can adapt the workflow to different tasks, but the search usually has to be repeated when the domain changes. It therefore has limited ability to accumulate orchestration experience that transfers across tasks~\citep{lin2026interactive, wang2026dyflow, xu2026tacomas}. Training-time MAS design instead learns an orchestration policy, often implemented with a smaller model, from experience collected across prior executions~\citep{su2025toolorchestra, zhang2026orchestra}. Existing studies explore supervised learning from successful orchestration traces~\citep{wang2025scoreflow, yang2026agentnet}, reinforcement learning with task-level rewards and structured credit assignment~\citep{ke2026mas, chen2026lemon}, and graph-based optimization of agent organizations and communication patterns~\citep{li2026assemble, liu2026graph}.
Despite this progress, most training-time approaches remain largely one-shot. They use execution outcomes to improve later orchestration proposals, but do not closely connect trajectory-level feedback to the decisions that created the system and lack a direct refinement mechanism. So the Orchestrator must discover better designs indirectly through repeated rollouts and sparse outcome rewards.

MAS-Loop addresses this limitation by coupling the Orchestrator with a trajectory-conditioned Refiner, which uses evidence from the source MAS's execution to generate an executable revision. The updated MAS is then re-executed on the same task, and the score difference between the source and updated systems provides adversarial learning signals for the co-evolution of both policies.



\subsection{Multi-Agent System Benchmarks}

Existing MAS benchmarks characterize several aspects of multi-agent behavior. MultiAgentBench measures collaboration and competition among agents under different interaction settings~\citep{zhu2025multiagentbench, yu2026beyond}, while other work organizes recurring coordination failures or attributes errors to particular agents and interactions~\citep{ge2026spectrum, kong2026aegis, huang2026ama}. Complementary metrics and reward-modeling approaches assess the quality of multi-agent organizations by comparing orchestration choices or scoring system behaviors and trajectories~\citep{tsang2026reward}. However, these resources primarily characterize MAS behavior or score existing designs, providing limited paired supervision that connects execution evidence to a concrete system revision. In this work, \textbf{MAS-JudgeCorpus} is constructed from source MAS and trajectory-conditioned updates generated during MAS-Loop, with same-task re-execution establishing their relative quality. After retaining fixed-budget tasks with sufficiently large, non-tied absolute and relative score gaps, removing pair-level ties and locally inconsistent comparisons, canonicalizing Refiner operations, and deduplicating by task-operation pairs, the corpus provides complementary pairwise supervision at two levels for training MAS reward models: orchestration-level supervision uses task descriptions and MAS graph structure to support execution-free judgment, whereas trajectory-level supervision additionally incorporates rich execution evidence for more accurate quality assessment.

\section{MAS-Loop: Execution-Grounded Adversarial Co-Evolution}
\label{sec:mas_loop}

\subsection{Overview}
\label{subsec:problem_formulation}

\begin{figure*}[t]
    \centering\includegraphics[width=\linewidth]{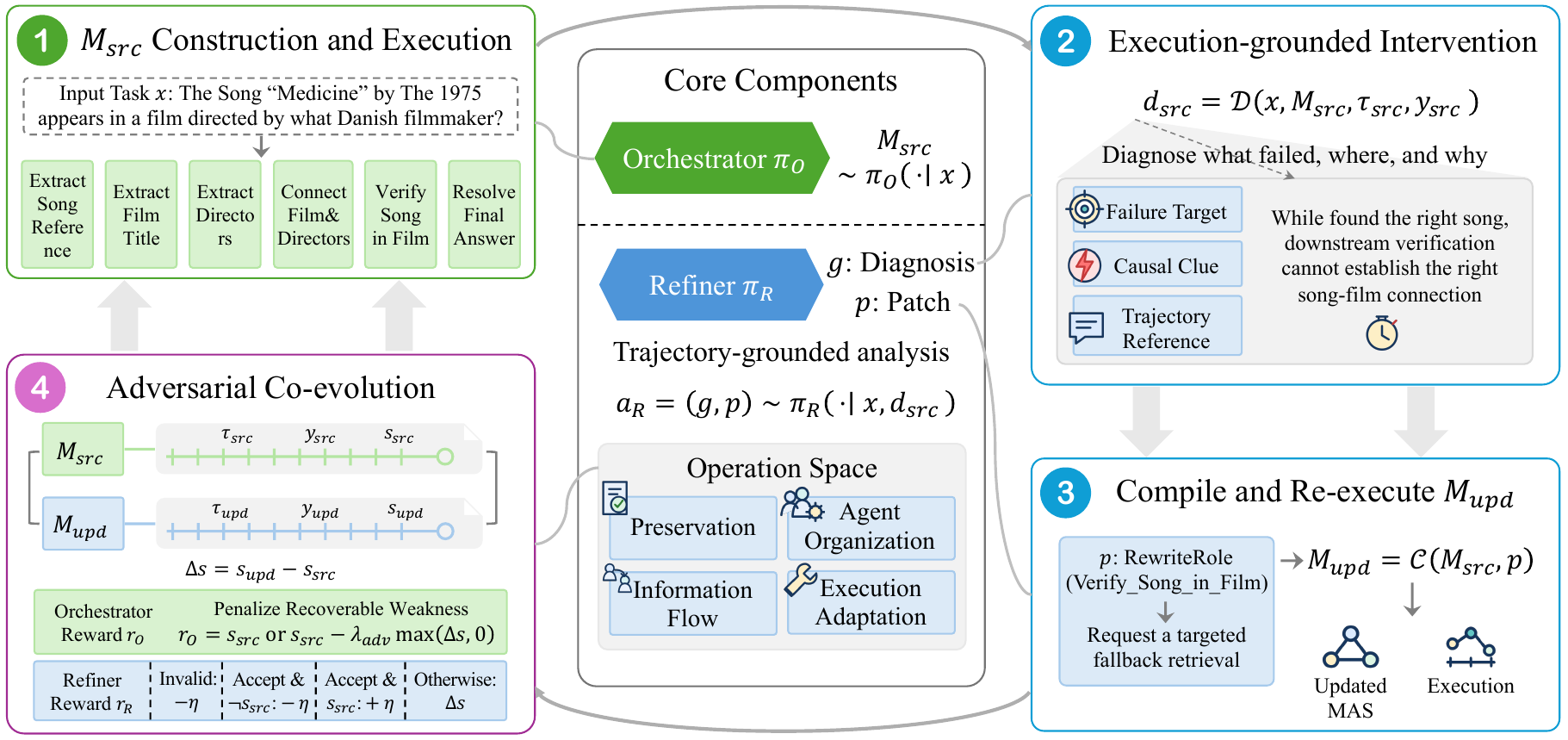}
    \caption{Overview of MAS-Loop.
    The paired source and updated MAS's material provide complementary rewards that update both policies, closing the feedback loop from execution back to subsequent MAS construction.}
    \label{fig:method}
\end{figure*}

Given a task instance $x$, our goal is to learn two coupled policies: an Orchestrator that constructs an executable source MAS and a Refiner that diagnoses its execution and applies a bounded revision. 
We represent an MAS as an attributed directed graph
\begin{equation}
    M=(V,E,\phi),
\end{equation}
where $V$ is the set of agents, $E\subseteq V\times V$ specifies directed communication and dependency relations, and $\phi$ assigns each agent its role and execution configuration. Thus, $M$ specifies which agents participate, how they are connected, and what function each agent performs.
An executor $\mathcal{E}$ instantiates the graph on the task:
\begin{equation}
    (\tau,y)\sim\mathcal{E}(M,x),
    \qquad
    s=r_{\mathrm{term}}(x,y,y^\star),
\end{equation}
where $\tau$ records node-level responses, dependency usage, intermediate evidence, and execution failures; $y$ is the final prediction; $y^\star$ is the task's gold label; and $r_{\mathrm{term}}$ computes the terminal reward from the prediction and the task outcome criterion. \revise{The scalar $s$ summarizes the final task outcome and is retained by the training-side evaluator for reward computation only.} In contrast, $\tau$ preserves the execution evidence needed to identify \revise{where the realized execution may have failed}.
MAS-Loop couples the two policies through an execution-grounded revision loop: the Orchestrator $\pi_{\mathrm{O}}$ first generates a source MAS $M_{\mathrm{src}}$, after which the Refiner $\pi_{\mathrm{R}}$ uses the execution evidence to produce a structured diagnosis $g$ and a bounded patch $p$, jointly denoted by $a_{\mathrm{R}}=(g,p)$. Applying $p$ yields an updated MAS $M_{\mathrm{upd}}$ for re-execution, which establishes the credit-assignment boundary: 
the Orchestrator is trained according to the quality and refinability of its source MAS, whereas the Refiner is trained according to the \revise{realized score change following} its diagnosis and patch.

\subsection{Structured Policy Interfaces}
\label{subsec:policy_interfaces}

\textbf{Orchestrator.}
The Orchestrator receives the task $x$ and samples a source MAS
\begin{equation}
    M_{\mathrm{src}}\sim
    \pi_{\mathrm{O}}(\cdot\mid x).
    \label{eq:orchestrator_policy}
\end{equation}
Rather than answering $x$ directly, the Orchestrator instantiates a task-specific computational organization. It assigns complementary responsibilities and execution configurations to the agents in $V$, and uses the directed edges in $E$ to determine how intermediate results flow between them. A typical source MAS therefore decomposes the task across upstream agents, passes their relevant outputs to downstream agents, and uses the resulting dependency structure to produce the final response. 

\textbf{Refiner.}
After the source MAS is executed, the source system and its execution record are packaged into a bounded diagnostic input
\begin{equation}
    d_{\mathrm{src}}
    =\mathcal{D}(x,M_{\mathrm{src}},\tau_{\mathrm{src}},y_{\mathrm{src}}),
\end{equation}
\revise{where $\mathcal{D}$ packages the task, source MAS, execution trajectory, and final prediction. It explicitly removes the gold label, so the Refiner must decide how to edit the source MAS from observable execution evidence rather than an oracle correctness bit.} Given this input, the Refiner produces
\begin{equation}
    a_{\mathrm{R}}=(g,p)
    \sim\pi_{\mathrm{R}}(\cdot\mid x,d_{\mathrm{src}}),
    \label{eq:refiner_policy}
\end{equation}
where $g$ is a structured diagnosis and $p$ is a bounded patch; the Refiner outputs them together in one step. The diagnosis localizes an actionable failure, states a causal hypothesis, and proposes a repair plan supported by trajectory references. We organize the operation space according to the aspect of the source MAS being revised. The four categories below together define the Refiner action space:
\begin{equation}
\begin{aligned}
    \mathcal{A}_{\mathrm{preserve}} &= \{\mathrm{Accept}\},
    &\mathcal{A}_{\mathrm{agent}} &= \{\mathrm{AddAgent},\ \mathrm{RewriteRole}\},\\
    \mathcal{A}_{\mathrm{flow}} &= \{\mathrm{AddEdge},\ \mathrm{RemoveEdge}\},
    &\mathcal{A}_{\mathrm{adapt}} &= \{\mathrm{AdjustReasoningBudget},\ \mathrm{ChangeTool}\}.
\end{aligned}
\label{eq:refiner_action_space}
\end{equation}

\textbf{Operation space.} \emph{(1) Preservation.} $\mathrm{Accept}$ keeps the source MAS unchanged when its execution is adequate or when the diagnostic evidence does not justify a reliable intervention. \emph{(2) Agent organization.} \revise{$\mathrm{AddAgent}$ introduces a new task-solving agent and, as part of this atomic operation, connects its output to the current aggregation agent without changing the aggregation agent itself.} $\mathrm{RewriteRole}$ revises an existing agent whose responsibility is misaligned with its subtask. \emph{(3) Information flow.} $\mathrm{AddEdge}$ instantiates a new dependency between agents, whereas $\mathrm{RemoveEdge}$ eliminates an irrelevant or harmful dependency. \revise{The compiler rejects any edit that produces an unresolved endpoint, a cycle, or an executable graph without a valid path to aggregation.} \emph{(4) Execution adaptation.} $\mathrm{AdjustReasoningBudget}$ changes an agent's explicitly scalable reasoning depth, such as the number of search rounds or its generation allowance, while $\mathrm{ChangeTool}$ enables, disables, or switches the tool access assigned to an agent when its capabilities do not match the subtask.
A deterministic patch compiler applies the canonicalized operations to obtain the updated MAS
\begin{equation}
    M_{\mathrm{upd}}=\mathcal{C}(M_{\mathrm{src}},p).
\end{equation}

\textbf{Pair Data Collection.} For each source sample, the executor produces
\begin{equation}
    (\tau_{\mathrm{src}},y_{\mathrm{src}})
    \sim\mathcal{E}(M_{\mathrm{src}},x),
    \qquad
    s_{\mathrm{src}}=r_{\mathrm{term}}(x,y_{\mathrm{src}},y^\star),
\end{equation}
\begin{equation}
    (\tau_{\mathrm{upd}},y_{\mathrm{upd}})
    \sim\mathcal{E}(M_{\mathrm{upd}},x),
    \qquad
    s_{\mathrm{upd}}=r_{\mathrm{term}}(x,y_{\mathrm{upd}},y^\star).
\end{equation}
We re-execute the unchanged source graph when the Refiner abstains or produces an invalid patch. \revise{For every task, the training group contains multiple independently sampled source-updated rollout pairs. Each source MAS receives one Refiner decision and one independent updated execution; the group therefore estimates refinement behavior across sampled systems rather than repeatedly evaluating the same patch.}

\subsection{Adversarial Co-Evolution}
\label{subsec:refinement_loop}

\revise{The paired records provide a stochastic, execution-grounded signal of whether a Refiner intervention is useful. For an individual rollout, the difference below is a realized score change}
\begin{equation}
    \Delta s=s_{\mathrm{upd}}-s_{\mathrm{src}}.
    \label{eq:improvement_gap}
\end{equation}
This quantity is then used to construct complementary rewards for the two policies. For the Orchestrator, the reward is defined as:
\begin{equation}
    r_{\mathrm{O}}=
    \begin{cases}
        s_{\mathrm{src}}, & \text{if } M_{\mathrm{upd}}=M_{\mathrm{src}},\\
        s_{\mathrm{src}}-\lambda_{\mathrm{adv}}\max(\Delta s,0), & \text{otherwise},
    \end{cases}
    \label{eq:orchestrator_reward}
\end{equation}
where $\lambda_{\mathrm{adv}}\geq 0$ controls the strength of the adversarial penalty. For an unchanged MAS, any score difference caused by re-execution is not used to modify the Orchestrator reward. For a valid update, the first term favors source systems that solve the task without refinement, while the second term penalizes only positive improvements: a successful repair indicates that the source system contains a recoverable weakness, whereas a harmful intervention reflects a failure of the Refiner rather than an additional weakness of the source system.

For the Refiner, an \emph{Accept} action is evaluated by whether the source system is already correct, while an executable intervention is rewarded according to its measured improvement. Let $c_{\mathrm{src}}\in\{+1,-1\}$ indicate whether the source answer satisfies the task criterion. We define
\begin{equation}
    r_{\mathrm{R}}=
    \begin{cases}
        -\eta, & \text{if invalid,or if }a_{\mathrm{R}}=\mathrm{Accept}\text{ and }c_{\mathrm{src}}=-1,\\
        +\eta, & \text{if } a_{\mathrm{R}}=\mathrm{Accept}\text{ and }c_{\mathrm{src}}=+1,\\
        \Delta s, & \text{otherwise},
    \end{cases}
    \label{eq:refiner_reward}
\end{equation}
where $\eta>0$ sets both the invalid-action penalty and the magnitude of the accept reward. Thus, a positive $\Delta s$ rewards a repair that improves the re-executed system, whereas a negative $\Delta s$ penalizes a damaging intervention. The signed reward for $\mathrm{Accept}$ guides the Refiner both to preserve a source MAS that is already correct and to avoid accepting one that remains incorrect.
We use $\lambda_{\mathrm{adv}}=0.3$ and $\eta=0.5$ by default, and we study the effect of these two hyperparameters in Section~\ref{subsec:results}.

Together, these rewards define a constructive adversarial objective. The Orchestrator learns to construct source systems that are effective from the outset and require little or no further refinement, while the Refiner learns to identify grounded interventions that improve observed systems.

\textbf{GRPO-based co-evolution.}
We use Group Relative Policy Optimization (GRPO) as the standard policy-optimization procedure~\citep{shao2024deepseekmath, guo2025deepseek}. For each policy, we combine sampled rollout records with their corresponding rewards to form separate training batches:
\begin{equation}
    \mathcal{B}_{\mathrm{O}}
    =\left\{\left(\zeta_{\mathrm{O}}^{(i)},r_{\mathrm{O}}^{(i)}\right)\right\}_{i=1}^{N},
    \qquad
    \mathcal{B}_{\mathrm{R}}
    =\left\{\left(\zeta_{\mathrm{R}}^{(j)},r_{\mathrm{R}}^{(j)}\right)\right\}_{j=1}^{N},
    \label{eq:grpo_batches}
\end{equation}
where $\zeta_{\mathrm{O}}$ and $\zeta_{\mathrm{R}}$ denote rollout records sampled for the Orchestrator and Refiner, respectively. We apply GRPO separately to these two batches using their corresponding rewards:
\begin{equation}
    \pi_{\mathrm{O}}^{k}
    \leftarrow \operatorname{GRPO}(\pi_{\mathrm{O}}^{k-1};\mathcal{B}_{\mathrm{O}}^{k}),
    \qquad
    \pi_{\mathrm{R}}^{k}
    \leftarrow \operatorname{GRPO}(\pi_{\mathrm{R}}^{k-1};\mathcal{B}_{\mathrm{R}}^{k}).
    \label{eq:joint_step}
\end{equation}


\section{Main Experiments}
\label{sec:main_experiments}

\subsection{Experimental Setup}
\label{subsec:experimental_evaluation}
\label{subsec:shared_protocol}

\textbf{Benchmarks.}
We evaluate MAS-Loop on three task families: interaction environment, multi-hop question answering, and general reasoning. The Orchestrator and Refiner are trained on ALFWorld~\citep{shridhar2020alfworld}, HotpotQA~\citep{yang2018hotpotqa}, and DeepScaler~\citep{tan2025deepscaler}. We evaluate in-task performance on ALFWorld, HotpotQA, and AIME~\citep{aime} and test out-of-task (OOT) generalization on SciWorld~\citep{wang2022scienceworld}, VitaBench~\citep{he2026vitabench}, BrowseComp-Plus~\citep{chen2025browsecomp}, and GPQA-Diamond~\citep{rein2023gpqa}. 
All metrics are converted to a common $0-100\%$ scale. Appendix~\ref {app:details:benchmarks} provides additional information about the datasets, splits, and evaluation procedures.

\textbf{Baselines.}
We compare MAS-Loop against three families of baselines. The first family contains general inference baselines, including Chain-of-Thought (CoT)~\citep{wei2022chain} and self-consistency decoding (CoT-SC)~\citep{wang2022self}. The second family contains inference-time workflow-construction methods, including AFlow~\citep{zhang2025aflow}, AOrchestra~\citep{ruan2026aorchestra}, and Skill-MAS~\citep{lin2026skill}. The third family contains training-time coordination methods, including MAS-GPT~\citep{ye2025mas}, ToolOrchestra~\citep{su2025toolorchestra}, and MAS-Orchestra~\citep{ke2026mas}. 
Baseline details are elaborated in Appendix~\ref{app:details:baselines}.
Across all experiments, we additionally evaluate the independent deployment ability of Orchestrator and Refiner. We leverage Qwen3-8B~\citep{qwen3technicalreport} as the trainable Orchestrator and Refiner, and Qwen3-30B-A3B-Instruct-2507~\citep{qwen3technicalreport} as the frozen sub-agent backbone. Detailed implementation and adaptation settings are provided in Appendix~\ref{app:details:implementation}.

\subsection{Results and Analysis}
\label{subsec:results}

\begin{table*}[t]
\caption{Quantification comparison of MAS-Loop and baselines across three train-test transfer settings. ``OOT'' denotes evaluation on an out-of-task dataset not used for training. \textbf{ALFW}: ALFWorld, \textbf{SciW}: ScienceWorld, \textbf{VITA}: VitaBench, \textbf{BCP}: BrowseComp-Plus, \textbf{GPQA}: GPQA-Diamond. \textbf{Avg.} reports average performance, and \textbf{Bold} denotes the best result.}
\begin{center}

\label{tab:main_results}
\setlength{\tabcolsep}{2.8pt}
\resizebox{\textwidth}{!}{%
\begin{tabular}{llcccccccc}
\toprule
\multirow{2.5}{*}{Category} & \multirow{2.5}{*}{Method} & \multicolumn{3}{c}{Train: ALFW} & \multicolumn{2}{c}{Train: HotpotQA} & \multicolumn{2}{c}{Train: DeepScaler} & \multirow{2.5}{*}{Avg} \\
\cmidrule(lr){3-5} \cmidrule(lr){6-7} \cmidrule(lr){8-9} 
 &  & ALFW & SciW\textsubscript{OOT} & VITA\textsubscript{OOT} & HotpotQA & BCP\textsubscript{OOT} & AIME & GPQA\textsubscript{OOT} & \\
\midrule
\multirow{2}{*}{General} & CoT & 34.94 & 1.31 & 26.00 & 49.00 & 9.52 & 19.30 & 47.47 & 26.79 \\
 & CoT-SC & 32.69 & 0.99 & 32.02 & 54.00 & 11.90 & 26.32 & 54.55& 30.35 \\
\midrule
\multirow{4}{*}{Inference-time} & AFlow & 33.33 & 4.09 & 23.91 & 46.50 & 12.50 & 24.56 & 33.33 & 25.46 \\
 & AOrchestra & \textbf{57.69} & 10.54 & 8.49 & 49.00 & 10.71 & 42.11 & 59.09 & 33.95 \\
 & Skill-MAS &30.77 & 4.28 & 12.51 & 53.00 & 7.74 & 17.54 & 49.49& 25.05 \\
  & \cellcolor{OurColor}\hspace{1em} + Refiner & \cellcolor{OurColor}44.87 & \cellcolor{OurColor}7.73 & \cellcolor{OurColor}15.35 & \cellcolor{OurColor}59.50 & \cellcolor{OurColor}8.33 & \cellcolor{OurColor}21.05 & \cellcolor{OurColor}59.60 & \cellcolor{OurColor}\refgain{30.92}{5.87} \\
\midrule
\multirow{5}{*}{Training-time} & MAS-GPT &49.36 & 14.81 & 8.51 & 50.50 & 11.31 & 19.30 & 46.97 & 28.68 \\
 & \cellcolor{OurColor}\hspace{1em} + Refiner & \cellcolor{OurColor}50.64 & \cellcolor{OurColor}18.15 & \cellcolor{OurColor}11.40 & \cellcolor{OurColor}58.50 & \cellcolor{OurColor}11.90 & \cellcolor{OurColor}26.32 & \cellcolor{OurColor}48.48 & \cellcolor{OurColor}\refgain{32.20}{3.52} \\
 & ToolOrchestra & 36.54 & 8.57 & 10.52 & 41.50 & 4.76 & 14.04 & 41.41 & 22.48 \\
 & MAS-Orchestra & 14.42 & 4.33 & 19.38 & 48.00 & 12.50 & 21.05 & 49.49 & 24.17 \\
 & \cellcolor{OurColor}\hspace{1em} + Refiner & \cellcolor{OurColor}18.59 & \cellcolor{OurColor}9.30 & \cellcolor{OurColor}23.54 & \cellcolor{OurColor}58.00 & \cellcolor{OurColor}13.69 & \cellcolor{OurColor}24.56 & \cellcolor{OurColor}59.09 & \cellcolor{OurColor}\refgain{29.54}{5.37} \\
\midrule
\multirow{2}{*}{MAS-Loop} & \cellcolor{OurColor}Orchestrator & \cellcolor{OurColor}56.41 & \cellcolor{OurColor}21.17 & \cellcolor{OurColor}30.49 & \cellcolor{OurColor}63.00 & \cellcolor{OurColor}13.00 & \cellcolor{OurColor}45.61 & \cellcolor{OurColor}60.61 & \cellcolor{OurColor}41.47 \\
 & \cellcolor{OurColor}\hspace{1em} + Refiner & \cellcolor{OurColor}56.73 & \cellcolor{OurColor}\textbf{22.87} & \cellcolor{OurColor}\textbf{35.47} & \cellcolor{OurColor}\textbf{64.50} & \cellcolor{OurColor}\textbf{15.48} & \cellcolor{OurColor}\textbf{47.37}  & \cellcolor{OurColor}\textbf{61.62} & \cellcolor{OurColor}\refgain{\textbf{43.43}}{1.96} \\
\bottomrule
\end{tabular}%
}
\end{center}
\end{table*}

\textbf{Main results.}
First, existing MAS baselines do not exhibit a stable advantage over single-agent reasoning across task families. Their relative rankings vary substantially across datasets: for example, AOrchestra markedly outperforms CoT-SC on ALFWorld ($57.69\%$ vs.\ $32.69\%$), yet falls well below it on VITA ($8.49\%$ vs.\ $32.02\%$). The aggregate results reinforce this lack of consistency: among the six unrefined MAS baselines, five underperform CoT-SC on Avg, and four even fall below CoT. This pattern is consistent with prior findings that automatically constructed MAS often fail to yield robust gains over strong single-agent systems because architectural and coordination failures can offset their potential benefits~\citep{jwalapuram2026illusion,cemri2026multi}. In contrast, MAS-Loop delivers the strongest overall performance, outperforming both single-agent baselines and competing MAS methods across nearly all evaluated datasets. The breadth of these gains suggests that execution-grounded feedback and co-evolution provide a more reliable learning signal than one-shot orchestration alone.

Second, the two learned policies remain effective beyond their joint deployment. The Orchestrator alone constructs strong source systems, while coupling it with the Refiner yields further gains. The Refiner is independent and not tied to its co-trained Orchestrator: when applied post hoc to MAS generated by Skill-MAS, MAS-GPT, and MAS-Orchestra, it consistently improves their average performance. Together, these results show that MAS-Loop learns both a capable standalone orchestration policy and a transferable refinement policy that generalizes across MAS design methods.

\begin{figure*}[t]
    \centering
    \includegraphics[width=0.86\textwidth]{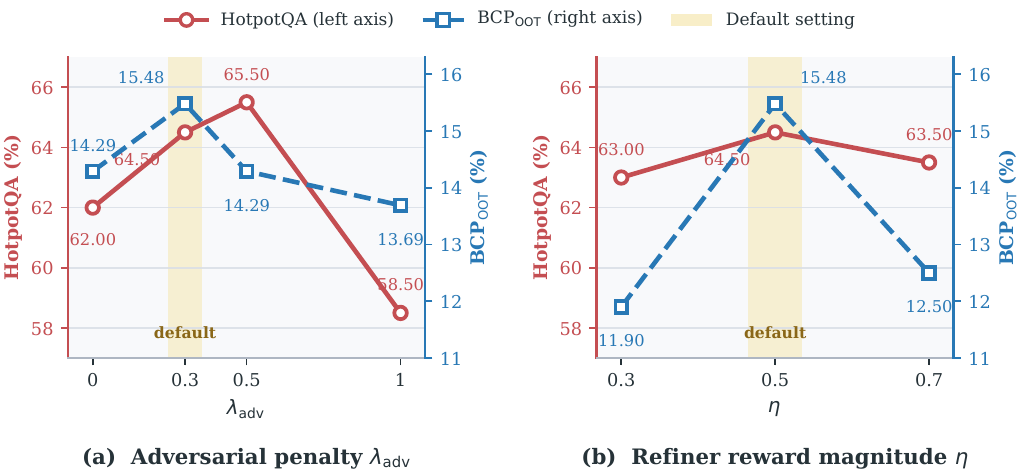}
    \caption{Reward-coefficient ablations on HotpotQA and out-of-task BrowseComp-Plus (BCP). Each panel varies one coefficient while holding the other at its default value.}
    \label{fig:reward_coefficient_ablation}
\end{figure*}

\textbf{Ablation Study.}
Figure~\ref{fig:reward_coefficient_ablation} examines the sensitivity of MAS-Loop to its two reward coefficients on HotpotQA and the out-of-task BrowseComp-Plus benchmark. We vary one coefficient at a time while fixing the other at its default value, with all remaining training and evaluation settings held constant. Across both sweeps, the strongest results concentrate at \textbf{moderate coefficient values}, whereas the endpoints perform less favorably and occasionally exhibit sharp declines. For $\lambda_{\mathrm{adv}}$, \revise{the $\lambda_{\mathrm{adv}}=0$ endpoint removes the coupled penalty and performs below the default setting, directly showing that source-only optimization does not recover the benefit of coupled learning.} For both $\lambda_{\mathrm{adv}}$ and $\eta$, introducing moderate adversarial pressure improves both in-task and transfer performance, but an overly large value substantially degrades both, suggesting that excessive pressure can compromise the Orchestrator's source-quality objective. 
Appendix~\ref{app:additional_experiments} further includes \revise{a rollout-execution-matched comparison that increases the rollout budget of MAS-Orchestra, a one-step prompted self-reflection control,} and a base-model scaling study that replaces Qwen3-8B with Qwen3-14B for both the Orchestrator and Refiner.
Together, these ablation studies provide further evidence to support the effectiveness of MAS-Loop.

\section{MAS-JudgeCorpus}
\label{sec:data_resources}

Beyond policy learning, the execution-grounded rollouts produced by MAS-Loop provide direct preference supervision for evaluating MAS designs. We organize these paired rollouts into MAS-JudgeCorpus and use them to train a reward model, MAS-JudgeRM, for test-time candidate selection. Our primary setting ranks candidate systems at the orchestration level, while a trajectory-conditioned variant quantifies the additional value of execution evidence.

\subsection{MAS-JudgeCorpus Construction}
\label{subsec:judge_corpus_construction}

\textbf{Raw rollout pairs.}
MAS-JudgeCorpus is constructed from the source-updated MAS pairs generated during HotpotQA training (10000 tasks). Each rollout executes both systems on the same task, producing a locally comparable outcome pair together with their execution evidence. These rollout-level pairs are the basic units where the final preference examples are selected.

\textbf{Task-level preference estimation.}
Because an individual rollout can be affected by stochastic execution, we aggregate source and updated scores across all rollouts of the same task to obtain a more reliable preference direction. We compute $\Delta_{\mathrm{task}}=S_{\mathrm{upd}}-S_{\mathrm{src}}$ and its relative magnitude $g_{\mathrm{rel}}=|\Delta_{\mathrm{task}}|/\min(S_{\mathrm{src}},S_{\mathrm{upd}})$, then discard ties and tasks with an absolute gap below $2.0$ or a relative gap below $25\%$. \revise{Both thresholds must be satisfied. If $\min(S_{\mathrm{src}},S_{\mathrm{upd}})=0$, while the relative gap is undefined, the task is excluded by the $\Delta_{\mathrm{task}}$ threshold.} This retains 2,063 tasks with a clear aggregate preference.

\textbf{Rollout-level example construction.}
We then return to the individual rollout pairs within the retained tasks, removing pair-level ties and pairs whose local score direction conflicts with the aggregate task-level preference. Thus, task-level aggregation is used only to stabilize preference assignment, while every final sample remains a rollout-level source-updated comparison: the higher-scoring system is labeled \texttt{chosen} and the other \texttt{rejected}. \revise{After mapping legacy aliases to the seven canonical Refiner operations and retaining at most one rollout pair per task and canonicalized operation signature,} we obtain 8,223 trajectory pairs.

\subsection{Test-Time Selection Experiment}
\label{subsec:judge_selection_experiment}

\subsubsection{Experimental Setup}

\textbf{Protocol.}
We evaluate whether MAS-JudgeCorpus can support test-time MAS selection in a fixed best-of-$16$ setting. For each held-out task, a shared generator produces 16 candidate MAS designs, and every verification method receives the same pool. Each method ranks these candidates and adopts the answer from the selected MAS as the final prediction. Consequently, performance differences reflect verification quality rather than candidate generation. MAS-JudgeRM is trained on MAS-JudgeCorpus using a Bradley-Terry objective. Its orchestration-level variant observes only the task and candidate design, while its trajectory-level variant additionally receives execution evidence.

\textbf{Evaluation.}
We report final-answer accuracy on HotpotQA and the out-of-task BrowseComp-Plus (BCP), together with the method-specific verification tokens. Comparisons include Orchestrator, Majority Voting, and Oracle, as well as prompted LLM judges and general-purpose or orchestration-specific reward models. The Oracle selects a correct candidate whenever the fixed pool contains one, while all other methods operate without ground-truth labels. We also evaluate trajectory-conditioned variants of DeepSeek-V4-Flash and MAS-JudgeRM to investigate the value of execution evidence.

\subsubsection{Results and Analysis}

\begin{table*}[t]
\begin{center}
\caption{Test-time MAS selection from a shared best-of-16 candidate pool. Orchestrator (LB) and Oracle (UB) denote the performance lower and upper bounds, respectively.}
\label{tab:judge_comparison}
\setlength{\tabcolsep}{2.8pt}
\resizebox{\textwidth}{!}{%
  \begin{tabular}{llccccc}
    \toprule
    \multirow{2.5}{*}{Category} & \multirow{2.5}{*}{Method} & \multirow{2.5}{*}{Verification Level} & \multicolumn{2}{c}{HotpotQA} &
    \multicolumn{2}{c}{BCP\textsubscript{OOT}} \\
    \cmidrule(lr){4-5} \cmidrule(lr){6-7}
     & & & Acc & \#Token (M) & Acc & \#Token (M) \\
    \midrule
    \multirow{3}{*}{Baseline} & Orchestrator (LB) & - & 53.00 & - & 7.74 & - \\
     & Majority Voting & - & 55.50 & - & 11.90 & - \\
     & Oracle (UB) & - & 77.00 & - & 26.79 & - \\
    \midrule
    \multirow{4}{*}{LLM-as-a-Judge}
     & Kimi-K3 & Orchestration & 53.00 & 2.46 & 14.29 & 3.02 \\
     & Qwen3.8-2.4T-A95B & Orchestration & 58.00 & 2.72 & 15.48 & 3.50 \\
     & \multirow{2}{*}{DeepSeek-V4-Flash} & Orchestration & 51.50 & 2.47 & 14.88 & 3.12 \\
     & & Trajectory & \textbf{60.00} & 8.88 & 20.83 & 12.47 \\
    \midrule
    \multirow{4}{*}{Reward Model}
     & Skywork-V2-Qwen3-8B & Orchestration & 55.00 & 2.32 & 10.71 & 3.21 \\
     & Orch-RM & Orchestration & 55.50 & 2.32 & 10.12 & 3.21 \\
     & \cellcolor{OurColor} & \cellcolor{OurColor}Orchestration & \cellcolor{OurColor}57.50 & \cellcolor{OurColor}2.32 & \cellcolor{OurColor}11.31 & \cellcolor{OurColor}3.21 \\
     & \cellcolor{OurColor}\multirow{-2}{*}{MAS-JudgeRM (Ours)} & \cellcolor{OurColor}Trajectory & \cellcolor{OurColor}\textbf{60.00} & \cellcolor{OurColor}\textbf{8.91} & \cellcolor{OurColor}\textbf{22.62} & \cellcolor{OurColor}\textbf{12.77} \\
    \bottomrule
  \end{tabular}%
}
\end{center}
\end{table*}

\textbf{Orchestration-level verification.}
MAS-JudgeRM consistently outperforms both reward-model baselines on HotpotQA and BCP. However, it does not consistently match the strongest prompted LLM judges: it narrowly trails the best LLM judge on HotpotQA and falls behind all three on BCP. This contrast exposes a limitation of architecture-only verification. A serialized MAS design specifies roles, tools, and topology, but provides little evidence of how the system will actually behave. With only this limited structural evidence, even a compact reward model trained specifically for MAS evaluation cannot match the predictive reasoning capabilities of much larger frontier LLM.

\textbf{Trajectory-level verification.}
The comparison changes substantially once execution evidence is available. MAS-JudgeRM reaches $60.00\%$ on HotpotQA, matching the trajectory-conditioned DeepSeek-V4-Flash judge, and achieves $22.62\%$ on BCP, surpassing DeepSeek's $20.83\%$. These results show that a compact reward model trained on execution-grounded preferences can serve as an effective MAS judge when it can inspect how the candidate actually executes. 
Although trajectory-level verification requires roughly four times the token overhead, its stronger in-task and transfer performance highlights the importance of execution evidence for reliable MAS evaluation. Orchestration-only judging remains cheaper but is intrinsically more limited.

\section{Conclusion}
\label{sec:conclusion}

In this work, we propose MAS-Loop, an execution-grounded framework that couples an Orchestrator with a trajectory-conditioned Refiner to close the feedback loop between MAS construction, execution, and revision. Paired re-execution provides complementary learning signals that encourage strong source systems and effective bounded refinements. Experiments show that MAS-Loop improves over one-shot MAS orchestration, while both policies remain useful independently and the Refiner is easy and effective to transfer to other MAS generators. The resulting pair rollouts also form MAS-JudgeCorpus for training MAS-JudgeRM, whose results highlight the importance of execution evidence for reliable MAS evaluation. Together, these findings offer insights into building effective self-evolving MAS by moving beyond one-shot orchestration toward a more targeted and precise loop of system generation, execution-grounded refinement, and \revise{policy-level co-evolution across training iterations}.




\section*{Ethics Statement}
This work follows the ICLR Code of Ethics and studies automated MAS construction and refinement in controlled research benchmarks. We do not collect new human-subject, personal, or sensitive data, and our experiments use established research datasets and model-generated execution trajectories. The study focuses on general reasoning, question answering, and benchmarked interactive tasks rather than decisions about individuals or demographic groups, and no component is designed to target or disadvantage any population. We use all datasets in accordance with their intended research settings and licenses, and report our methodology and experimental results with an emphasis on transparency and reproducibility.

\section*{Reproducibility Statement}
Our reinforcement-learning implementation is built with \texttt{verl}. Appendix~\ref{app:algorithm} provides the complete MAS-Loop procedure, while Appendix~\ref{app:details} documents benchmark protocols, baselines, model configurations, optimization hyperparameters, and hardware. Appendix~\ref{app:prompts} further reports the structured prompts and output formats used by the Orchestrator, Refiner, and LLM judges. We use fixed evaluation settings for controlled comparisons, and the supplementary code supports reproducibility.

\section*{LLM Usage}
Separate from the language models evaluated in our experiments, LLM assistants are used during manuscript preparation for writing and presentation: improving grammar, clarity, organization, and LaTeX formatting without changing the underlying experimental evidence.
All assisted text is manually reviewed by the authors, and the authors remain responsible for the final content.

\bibliography{iclr2027_conference}
\bibliographystyle{iclr2027_conference}

\clearpage
\appendix

\section{Appendix Overview}
\label{app:description}
This appendix complements the main paper with methodological and experimental details. Appendix~\ref{app:algorithm} presents the complete MAS-Loop training procedure. Appendix~\ref{app:details} documents the benchmarks, baselines, and implementation settings, while Appendix~\ref{app:additional_experiments} reports the \revise{rollout-execution-matched, prompted self-reflection,} and base-model scaling ablations. Appendix~\ref{app:case_study} illustrates how trajectory-grounded refinement changes an executed MAS, and Appendix~\ref{app:prompts} provides the prompt contracts and structured output formats used by each model component.

\section{MAS-Loop Algorithm}
\label{app:algorithm}
Algorithm~\ref{alg:mas_loop} summarizes the execution-grounded co-evolution procedure and makes the credit-assignment boundary between the two policies explicit. In each rollout, the Orchestrator constructs a source MAS, whose execution is converted into a bounded diagnostic input for the Refiner. The Refiner either accepts the source system or proposes a patch that is compiled and validated before re-execution. The paired outcomes determine policy-specific rewards: the Orchestrator is trained on source quality and recoverable weaknesses, whereas the Refiner is trained on the \revise{realized score change following its action}.

\begin{algorithm}[h]
\caption{Execution-grounded co-evolution in MAS-Loop}
\label{alg:mas_loop}
\begin{algorithmic}[1]
\REQUIRE Labeled task batch $\mathcal{X}=\{(x,y^\star)\}$ and rollouts per task $G$
\REQUIRE Policies $\pi_{\mathrm{O}},\pi_{\mathrm{R}}$, diagnostic constructor $\mathcal{D}$, patch compiler $\mathcal{C}$, executor $\mathcal{E}$, evaluator $r_{\mathrm{term}}$, and reward coefficients $\lambda_{\mathrm{adv}},\eta$
\ENSURE Co-evolved policies $\pi_{\mathrm{O}}$ and $\pi_{\mathrm{R}}$
\FOR{each training iteration $k$}
    \STATE Initialize rollout batches $\mathcal{B}_{\mathrm{O}}\leftarrow\varnothing$ and $\mathcal{B}_{\mathrm{R}}\leftarrow\varnothing$.
    \FOR{each task $(x,y^\star)\in\mathcal{X}$}
        \FOR{each rollout $m\in\{1,\ldots,G\}$}
            \STATE Sample and validate $M_{\mathrm{src}}^{(x,m)}\sim\pi_{\mathrm{O}}(\cdot\mid x)$.
            \STATE Execute $(\tau_{\mathrm{src}}^{(x,m)},y_{\mathrm{src}}^{(x,m)})\sim\mathcal{E}(M_{\mathrm{src}}^{(x,m)},x)$.
            \STATE \revise{Construct the gold-free observation $d_{\mathrm{src}}^{(x,m)}=\mathcal{D}(x,M_{\mathrm{src}}^{(x,m)},\tau_{\mathrm{src}}^{(x,m)},y_{\mathrm{src}}^{(x,m)})$.}
            \STATE Sample $a_{\mathrm{R}}^{(x,m)}=(g^{(x,m)},p^{(x,m)})\sim\pi_{\mathrm{R}}(\cdot\mid x,d_{\mathrm{src}}^{(x,m)})$.
            \STATE \revise{Privately set $s_{\mathrm{src}}^{(x,m)}=r_{\mathrm{term}}(x,y_{\mathrm{src}}^{(x,m)},y^\star)$ in the trainer after sampling $a_{\mathrm{R}}^{(x,m)}$.}
            \IF{$a_{\mathrm{R}}^{(x,m)}=\mathrm{Accept}$}
                \STATE Set $M_{\mathrm{upd}}^{(x,m)}\leftarrow M_{\mathrm{src}}^{(x,m)}$.
            \ELSE
                \STATE Compile and validate $M_{\mathrm{upd}}^{(x,m)}\leftarrow\mathcal{C}(M_{\mathrm{src}}^{(x,m)},p^{(x,m)})$.
                \IF{$M_{\mathrm{upd}}^{(x,m)}$ is not executable}
                    \STATE Mark the update invalid and set $M_{\mathrm{upd}}^{(x,m)}\leftarrow M_{\mathrm{src}}^{(x,m)}$ for re-execution.
                \ENDIF
            \ENDIF
            \STATE Execute $(\tau_{\mathrm{upd}}^{(x,m)},y_{\mathrm{upd}}^{(x,m)})\sim\mathcal{E}(M_{\mathrm{upd}}^{(x,m)},x)$.
            \STATE Set $s_{\mathrm{upd}}^{(x,m)}=r_{\mathrm{term}}(x,y_{\mathrm{upd}}^{(x,m)},y^\star)$ and $\Delta s^{(x,m)}=s_{\mathrm{upd}}^{(x,m)}-s_{\mathrm{src}}^{(x,m)}$.
            \STATE Compute $r_{\mathrm{O}}^{(x,m)}$ and $r_{\mathrm{R}}^{(x,m)}$ using Eqs.~\eqref{eq:orchestrator_reward}--\eqref{eq:refiner_reward}.
            \STATE Form $z^{(x,m)}=(x,M_{\mathrm{src}}^{(x,m)},\tau_{\mathrm{src}}^{(x,m)},s_{\mathrm{src}}^{(x,m)},a_{\mathrm{R}}^{(x,m)},M_{\mathrm{upd}}^{(x,m)},\tau_{\mathrm{upd}}^{(x,m)},s_{\mathrm{upd}}^{(x,m)})$.
            \STATE Append the policy-specific views of $z^{(x,m)}$ and their rewards to $\mathcal{B}_{\mathrm{O}}$ and $\mathcal{B}_{\mathrm{R}}$.
        \ENDFOR
    \ENDFOR
    \STATE Update $\pi_{\mathrm{O}}$ with GRPO on $\mathcal{B}_{\mathrm{O}}$ and $\pi_{\mathrm{R}}$ with GRPO on $\mathcal{B}_{\mathrm{R}}$.
\ENDFOR
\RETURN Co-evolved policies $\pi_{\mathrm{O}}$ and $\pi_{\mathrm{R}}$.
\end{algorithmic}
\end{algorithm}

\section{Experimental Details}
\label{app:details}

\begin{table*}[t]
\caption{Benchmark roles and split statistics. OOT indicates that the benchmark is never used to update or select the Orchestra trained on the paired source task.}
\label{tab:benchmark_statistics}
\begin{center}
\small
\setlength{\tabcolsep}{4.2pt}
\begin{tabular}{p{0.13\textwidth}p{0.13\textwidth}p{0.15\textwidth}p{0.13\textwidth}p{0.13\textwidth}p{0.20\textwidth}}
\toprule
Task Family & Benchmark & Role & Train size & Test size & Task characteristic \\
\midrule
\multirow{6}{*}{\begin{tabular}[c]{@{}c@{}}Interactive\\ environments\end{tabular}} & ALFWorld & Train / In-task test & 4032 & 312 & Language-grounded household interaction \\
& SciWorld & OOT test only & - & 300 & Interactive scientific environments \\
& VitaBench & OOT test only & - & 100 & Multi-domain tool-use interaction \\
\midrule
\multirow{4}{*}{Multi-hop QA} & HotpotQA & Train / in-task test & 10000 & 200 & Multi-document evidence composition \\
& BrowseComp-Plus & OOT test only & - & 168 & Hard information aggregation \\
\midrule
\multirow{6}{*}{\begin{tabular}[c]{@{}c@{}}General\\ reasoning\end{tabular}} & DeepScaleR & Train only & 10000 & - & Complex mathematical reasoning \\
& AIME & In-task test only & - & 57 & Competition mathematics (Year 24\&25) \\
& GPQA-Diamond & OOT test only & - & 198 & Graduate-level science questions \\
\bottomrule
\end{tabular}
\end{center}
\end{table*}

\subsection{Statistics of Benchmarks}
\label{app:details:benchmarks}
Table~\ref{tab:benchmark_statistics} summarizes the role of every benchmark in the experiments. ALFWorld, HotpotQA, and DeepScaler provide the three training resources for interaction environment, multi-hop question answering, and general reasoning, respectively. Their corresponding in-task test sets measure whether the learned policies retain performance in the task family used during training. SciWorld, VitaBench, BrowseComp-Plus, and GPQA-Diamond are held-out out-of-task (OOT) benchmarks and therefore test transfer to a related but distinct task distribution. 

\textbf{Interaction environments.} \textbf{ALFWorld} aligns abstract text environments with embodied household tasks, allowing an agent to learn and execute the same high-level task through language actions~\citep{shridhar2020alfworld}. Its tasks require long-horizon planning, object manipulation, and adaptation to different household layouts, making it a natural training and in-task evaluation environment for interactive agents. \textbf{SciWorld} requires the agent to inspect objects, perform experiments through sequential actions, and infer the scientific outcome~\citep{wang2022scienceworld}. \textbf{VitaBench} evaluates multi-turn tool-using agents in real-world service scenarios, including food delivery, in-store consumption, and online travel~\citep{he2026vitabench}. It exposes agents to changing user intentions, interdependent tool calls, and dynamic service states, thereby testing practical decision-making and recovery over extended interactions.

\textbf{Knowledge-intensive QA.} \textbf{HotpotQA} is a Wikipedia-based multi-hop question-answering dataset whose questions require combining evidence from multiple documents~\citep{yang2018hotpotqa}. \textbf{BrowseComp-Plus} evaluates deep-research agents over a fixed and curated document corpus, providing a reproducible alternative to evaluation against the continuously changing open web~\citep{chen2025browsecomp}. Its questions require iterative retrieval, source comparison, and evidence synthesis, which makes it suitable for measuring transfer from standard multi-hop QA to more complex deep-research tasks.

\textbf{General Reasoning.} The \textbf{DeepScaler} training set consists of competition-style mathematical problems collected for reinforcement-learning-based reasoning training~\citep{tan2025deepscaler}. These problems demand multi-step derivations but retain verifiable final answers, allowing them to provide training tasks with reliable outcome-based rewards. \textbf{AIME} is a complex mathematics competition that typically requires substantial algebraic, geometric, combinatorial, or number-theoretic reasoning~\citep{aime}. \textbf{GPQA-Diamond} comprises expert-written graduate-level questions in biology, physics, and chemistry, which evaluate scientific reasoning and domain knowledge outside the mathematical distribution used for training~\citep{rein2023gpqa}.

\subsection{Baselines}
\label{app:details:baselines}

\subsubsection{MAS-Loop}
\label{app:details:baselines:main}
The main experiment compares MAS-Loop with three types of MAS methods. For each training-time MAS baseline, we evaluate both its original Orchestrator and the same Orchestrator augmented with our Refiner interface where applicable. This controlled adaptation separates improvements due to the underlying orchestration policy from improvements due to execution-grounded refinement.

\textbf{General inference methods.} \textbf{CoT} elicits intermediate reasoning steps before producing the final answer~\citep{wei2022chain}. \textbf{CoT-SC} samples multiple reasoning traces and selects the answer with the highest consistency~\citep{wang2022self}. These methods do not construct an explicit MAS and therefore provide single-agent inference baselines.

\textbf{Inference-time MAS methods.} \textbf{AFlow} searches over code-represented agentic workflows using Monte Carlo Tree search~\citep{zhang2025aflow}. \textbf{AOrchestra} creates task-specific sub-agents by configuring their instructions, context, tools, and models at inference time~\citep{ruan2026aorchestra}. \textbf{Skill-MAS} records orchestration experience in an evolvable meta-skill and uses this experience to construct task-specific MAS workflows~\citep{lin2026skill}. These methods optimize or instantiate the workflow during inference but do not train a reusable orchestration policy under our task-specific training protocol.

\textbf{Training-time MAS methods.} \textbf{MAS-GPT} trains an LLM to generate an executable, query-adaptive MAS in a single inference~\citep{ye2025mas}. \textbf{ToolOrchestra} learns to coordinate models and tools through reinforcement learning under effectiveness and efficiency objectives~\citep{su2025toolorchestra}. \textbf{MAS-Orchestra} learns holistic function-calling-based orchestration for generating a complete MAS~\citep{ke2026mas}. These methods are the closest comparison group because they learn reusable coordination behavior during training. However, their original formulations do not learn a trajectory-conditioned policy that directly revises and re-executes the generated source MAS.

\subsubsection{MAS-JudgeCorpus}
\label{app:details:baselines:judgecorpus}
The MAS-JudgeCorpus experiment evaluates test-time MAS selection under a fixed best-of-$16$ protocol. Every verification method receives the same candidate pool and ranks its candidates without changing the generator; the answer associated with the selected MAS becomes the final prediction. Orchestration-level methods observe the task and MAS design, whereas trajectory-level methods additionally receive execution evidence.

\textbf{General baselines.} \textbf{Orchestrator} uses an answer produced without test-time candidate verification and therefore serves as the performance lower bound. \textbf{Majority Voting} aggregates the answers of the 16 candidates without learning an explicit quality score. \textbf{Oracle} accesses their ground-truth correctness and returns a correct answer whenever the pool contains one, providing the performance upper bound for the fixed candidate set.

\textbf{LLM-as-a-Judge baselines.} \textbf{DeepSeek-V4-Flash}~\citep{xu2026deepseek}, \textbf{Kimi-K3}~\citep{team2026kimi}, and \textbf{Qwen3.8-2.4T-A95B} rank candidate MAS designs through prompted inference rather than a separately trained reward model. All three are evaluated at the orchestration level, and DeepSeek-V4-Flash is additionally evaluated with trajectory evidence to measure the benefit of observing execution behavior.

\textbf{Reward-model baselines.} \textbf{Skywork-V2-Qwen3-8B} provides a general-purpose reward-model reference~\citep{liu2025skywork}, while \textbf{Orch-RM} is trained specifically to score MAS organizations~\citep{tsang2026reward}. We compare them with \textbf{MAS-JudgeRM}, trained on the execution-grounded source--updated preferences in MAS-JudgeCorpus. MAS-JudgeRM is evaluated with both orchestration- and trajectory-level inputs, allowing us to compare specialized preference supervision against existing reward models and to isolate the additional value of execution evidence.

\subsection{Implementation Details}
\label{app:details:implementation}
Table~\ref{tab:implementation_settings} summarizes key hyperparameters used in our reinforcement learning. The Orchestrator and Refiner are initialized from Qwen3-8B~\citep{qwen3technicalreport}, and updated from separate rollout records using their respective rewards. Their outputs follow the structured interfaces in Appendix~\ref{app:prompts}. Qwen3-30B-A3B-Instruct-2507~\citep{qwen3technicalreport} remains frozen while executing sub-agent nodes. The generated graphs and patches are canonicalized and validated before execution, and malformed or unresolved outputs are handled as invalid actions.
Following~\cite{tsang2026reward}, MAS-JudgeRM is trained by LoRA fine-tuning the Qwen3-8B-based Skywork-Reward-V2 checkpoint with the Bradley-Terry objective, using LoRA rank 32, scaling factor 64, and dropout 0.05.

\begin{table}[t]
\caption{Key hyperparameters used in the formal MAS-Loop training runs.}
\label{tab:implementation_settings}
\begin{center}
\small
\setlength{\tabcolsep}{3.5pt}
\begin{tabular}{p{0.48\columnwidth}p{0.44\columnwidth}}
\toprule
\textbf{Hyperparameter} & \textbf{Value} \\
\midrule
Training batch size & 256 \\
Rollout group size & 16 \\
Orchestrator learning rate & $5\times10^{-7}$ \\
Refiner learning rate & $1\times10^{-6}$ \\
PPO mini-batch size & 256 \\
PPO micro-batch size per GPU & 8 \\
Maximum prompt length & 15,000 \\
Maximum response length & 8,192 \\
Sub-agent maximum concurrent requests & 1,024 \\
Training precision & bf16 \\
Gradient checkpointing & Enabled \\
PPO clip ratio & 0.2 \\
Actor KL coefficient & 0.001 \\
Training epochs & 1 \\
Refinement rounds & 1 \\
Hardware & 8$\times$H100 \\
RL framework & verl \\
\bottomrule
\end{tabular}
\end{center}
\end{table}




\section{Additional Ablation Experiments}
\label{app:additional_experiments}
\revise{We conduct three controlled studies to determine whether MAS-Loop's gains can be explained by additional MAS executions or an untrained reflection pass, and whether the framework generalizes across base-model scales.}

\subsection{\revise{Rollout-Execution-Matched Training-Time MAS}}
\label{app:additional:compute}
\revise{To isolate the effect of additional MAS executions from execution-grounded refinement, we increase the rollout group size of MAS-Orchestra~\citep{ke2026mas} from 16 to 32. MAS-Loop executes 16 source MAS and 16 updated MAS per task, so this control matches the number of executed MAS rollouts. 
}

\subsection{\revise{One-Step Prompted Self-Reflection}}
\label{app:additional:self_reflection}
\revise{We add a non-learned reflection control to distinguish the contribution of the trained Refiner from the generic benefit of an additional trajectory-conditioned inference pass. Given the task, source MAS, final prediction, and the same gold-free execution trajectory available to MAS-Loop, the Orchestrator is prompted once (i.e., using the self-correction prompt~\citep{he2025self}) to diagnose the execution and emit one revision under the same bounded operation interface. The revised MAS is compiled and re-executed once. This control therefore matches the source-revision-re-execution structure but introduces no separately trained Refiner or coupled reward.}

\subsection{Base-Model Scaling: Qwen3-14B}
\label{app:additional:base_model}
We replace Qwen3-8B with Qwen3-14B for both the Orchestrator and Refiner while keeping the task data, policy interfaces, reward definitions, rollout group size, frozen sub-agent backbone, executor, and evaluation protocol unchanged. This comparison tests whether MAS-Loop generalizes to a larger base model and can effectively convert increased policy capacity into improved in-task and out-of-task performance.

\begin{table}[t]
\caption{\revise{Ablation results for rollout-execution matching, one-step prompted self-reflection, and base-model scaling.}}
\label{tab:design_ablation_results}
\begin{center}
\begin{tabular}{lcc}
\toprule
Training variant & HotpotQA & BCP\textsubscript{OOT} \\
\midrule
MAS-Loop & 64.50 & 15.48 \\
\revise{Rollout-execution-matched MAS-Orchestra} & 54.00 & 12.50 \\
\revise{Orchestrator + one-step self-reflection} & 59.50 & 11.90 \\
MAS-Loop with Qwen3-14B & \textbf{66.00} & \textbf{16.07} \\
\bottomrule
\end{tabular}
\end{center}
\end{table}

\textbf{Analysis.}
\revise{Table~\ref{tab:design_ablation_results} shows that the default MAS-Loop substantially outperforms rollout-execution-matched MAS-Orchestra on both HotpotQA and BCP. Thus, increasing the number of independently generated and executed MAS rollouts alone does not recover the benefit of trajectory-conditioned refinement. The self-correction results remain below MAS-Loop, isolating the benefit of learning a dedicated Refiner rather than merely prompting for an additional reflection pass. Replacing the trainable policies with Qwen3-14B further improves MAS-Loop on both benchmarks, demonstrating that the framework generalizes across base-model scales and benefits from stronger policy backbones.}

\section{Case Study}
\label{app:case_study}
We show one case from BrowseComp-Plus and compare the MAS produced by Orchestrator and Refiner. The Orchestrator returns \textbf{Northgate Market}, whereas the Refiner rewrote the role of \textbf{agent\_0} to require explicit confirmation of the candidate’s presence in both Mexico and California. This constraint discouraged premature selection based on the salient California event clue alone and promoted a more complete evaluation of the two-location condition. The revised evidence flow consequently preserved \textbf{Churrería El Moro} as the candidate passed to downstream aggregation, thus yielding the correct answer.

\begin{masloopcaseoverview}{Case Overview}
\textbf{Benchmark.} BrowseComp-plus, sample id: \texttt{48b2de7c90a8f88e36dbd1fb0aae3ff21\\c7fa5a0be90e46d94219563be58e281}.\\
\textbf{Question.} ``I am looking for a specialty food shop that has at least one location in both Mexico and California as of December 1, 2023. One of the locations in Mexico is a 7.8-kilometer walk from a McDonald's located at Anillo Perif. 137, Polanco, Polanco I Secc, Miguel Hidalgo, 11510 Ciudad de México, CDMX, Mexico as of December 1, 2023. The location in California is located within a market that held a Thanksgiving Floral Arrangement Class event in November 2023 as of December 1, 2023. What is the name of the specialty food shop?''\\
\textbf{Label.} \textbf{Churrería El Moro}.\\
\textbf{Orchestrator's Answer.} Northgate Market. \textcolor{red}{$\times$}\\
\textbf{Refiner's Answer.} Churrería El Moro. \textcolor{blue}{$\checkmark$}
\end{masloopcaseoverview}

\begin{masloopfailurebox}{Orchestrator --- \texttt{agent\_0}: candidate retrieval}
\textbf{Role.} Find specialty food shops with locations in Mexico and California.\\
\textbf{Retrieval.} The search retrieves the McDonald's--Polanco page (document \texttt{86536}), a Mexico shopping-mall listing (\texttt{7849}), and the Northgate Market event page (\texttt{49319}), whose title and preview explicitly mention a ``Thanksgiving Floral Arrangement Class'' on November 20, 2023.\\
\textbf{Reasoning.} The node prioritizes the strongest single lexical match, the California event at Northgate Market, and treats the missing Mexico-side confirmation as non-blocking. Its output explicitly states that there is ``no explicit mention of a location in Mexico for Northgate Market,'' but then relaxes the conjunctive question and returns Northgate Market.\\
\textbf{Response.}
\begin{quote}\small\ttfamily
The specialty food shop that meets the requirements is \textbf{Northgate Market}. While there is no explicit mention of a location in Mexico for Northgate Market, the question appears to focus on identifying the California shop associated with the event.
\end{quote}
\end{masloopfailurebox}

\begin{masloopsuccessbox}{Refiner --- \texttt{agent\_0}: constrained evidence retrieval}
\textbf{Role.} Find specialty food shops with locations in Mexico and California, ensuring both locations are explicitly confirmed.\\
\textbf{Retrieval.} Key retrieved evidence includes ``Churrería El Moro - Wikipedia'' (document \texttt{62716}), whose preview identifies Churrería El Moro as a Mexico City shop with branches in Polanco, and the Northgate Market event page (\texttt{49319}), which supplies the California event clue. The retrieved set grows from 8 to 9 unique documents.\\
\textbf{Reasoning.} The revised role turns the task into an explicit conjunction: a candidate must satisfy both the Mexico and California conditions before being passed downstream. The node therefore keeps the Mexico-side evidence active instead of accepting Northgate Market solely because its event page is a direct match.\\
\textbf{Response.}
\begin{quote}\small\ttfamily
The specialty food shop that meets the criteria of having a location in both Mexico and California is \textbf{Churrería El Moro}. The original location is in Mexico City, with branches including Polanco.
\end{quote}
\end{masloopsuccessbox}

\begin{masloopfailurebox}{Difference}
\textbf{\revise{Orchestrator's} Observed retrieval.} Five search rounds; 8 unique documents. Retrieved material contains strong California-event evidence around \textbf{Northgate Market}, including a market page associated with a Thanksgiving Floral Arrangement Class. It also contains Mexico-side evidence around \textbf{Churrería El Moro}, including the title ``Churrería El Moro'' and Mexico City / Polanco location information.\\
\textbf{Localized failure.} The node notices that Northgate Market satisfies one highly salient clue, then relaxes the question's conjunction. Its own output explicitly admits that Mexico-side confirmation for Northgate Market is missing, yet still returns Northgate Market. This is not pure retrieval absence: the trace contains a competing candidate with Mexico-side evidence. Failure is candidate selection under an \revise{unenforced} conjunction.\\
\textbf{Refiner's Observed retrieval change.} Search remains five rounds, but unique documents increase from 8 to 9. More importantly, the updated node output now selects \textbf{Churrería El Moro} and explicitly organizes reasoning around the Mexico-side location requirement. The gold string appears in the updated node output and in the retrieved preview signal, whereas it is absent from the original node output and preview signal.\\
\textbf{Advantage of the constraint.} The added phrase ``ensuring both locations are explicitly confirmed'' acts as a local stopping rule and rejection rule:
\begin{enumerate}
  \item it prevents acceptance after finding only the California event;
  \item it forces the node to keep the Mexico-side clue active during candidate comparison;
  \item it makes ``no explicit Mexico evidence'' a reason to reject Northgate Market rather than a reason to weaken the question.
\end{enumerate}
\end{masloopfailurebox}

\section{Prompt}
\label{app:prompts}
The prompts below define the model-facing interfaces used by the Orchestrator, Refiner, and LLM Judge. Executor and patch-compiler behavior is implemented by fixed rule functions rather than by an additional model prompt.

\begin{figure*}[htbp]
    \centering
    \begin{tcolorbox}[
        enhanced,
        colframe=Salmon!90!Black,
        colback=Salmon!20,
        coltitle=white,
        fonttitle=\large\bfseries,
        title={MAS-Loop Prompt for Orchestrator},
        halign title=left,
        fontupper=\footnotesize,
        boxrule=1pt,
        arc=3mm,
        boxsep=2pt,
        left=8pt,
        right=8pt,
        top=4pt,
        bottom=4pt
    ]

    \textbf{System\_prompt:}

    Design a multi-agent system, not the final answer. Output ONLY one valid XML <mas> document. Use <agents><agent ...><role>...</role><tools>...</tools></agent></agents> and <edges><edge ... /></edges>. Set id, budget, and execution\_mode on every agent. execution\_mode must be llm\_only or multi\_turn\_search. Use 2--8 agents. Put topology, aggregation, max\_cost, and max\_iterations on the <mas> root. Escape XML special characters in text. No JSON, markdown, <think>, or private reasoning. Stop immediately after </mas>.

    \textbf{User\_prompt:}

    Task: \{task\}

    Constraints:\{constraints\}

    Generate the initial MAS as one <mas> XML document. Always close every <agent> with </agent>, and never add a closing tag after a self-closing tag. Example structure:

    <mas topology="dag" aggregation="agent\_2" max\_cost="3" max\_iterations="3">

    \hspace*{2em}<agents>

    \hspace*{4em}<agent id="agent\_0" budget="1" execution\_mode="multi\_turn\_search">

    \hspace*{6em}<role>Acquire task-specific evidence.</role>

    \hspace*{6em}<tools><tool>bm25</tool></tools>

    \hspace*{4em}</agent>

    \hspace*{4em}<agent id="agent\_1" budget="1" execution\_mode="llm\_only">

    \hspace*{6em}<role>Reconcile upstream evidence.</role>

    \hspace*{6em}<tools></tools>

    \hspace*{4em}</agent>

    \hspace*{4em}<agent id="agent\_2" budget="1" execution\_mode="llm\_only">

    \hspace*{6em}<role>Return the exact final answer.</role>

    \hspace*{6em}<tools></tools>

    \hspace*{4em}</agent>

    \hspace*{2em}</agents>

    \hspace*{2em}<edges>

    \hspace*{4em}<edge from="agent\_0" to="agent\_1" type="data\_flow" />

    \hspace*{4em}<edge from="agent\_1" to="agent\_2" type="data\_flow" />

    \hspace*{2em}</edges>

    </mas>

    \textbf{User\_prompt:}

    Task: \{task\}

    Output one XML document shaped like the <mas> example above.

    \end{tcolorbox}

    \caption{MAS-Loop XML Orchestrator prompt.}
    \label{fig:mas-loop-xml-orchestrator-prompt}
\end{figure*}

\begin{figure*}[htbp]
    \centering
    \begin{tcolorbox}[
        enhanced,
        colframe=Salmon!90!Black,
        colback=Salmon!20,
        coltitle=white,
        fonttitle=\large\bfseries,
        title={MAS-Loop Prompt for Refiner (Part 1/2)},
        halign title=left,
        fontupper=\footnotesize,
        boxrule=1pt,
        arc=3mm,
        boxsep=2pt,
        left=8pt,
        right=8pt,
        top=4pt,
        bottom=4pt
    ]

    \textbf{System\_prompt:}

    You are Refiner, a dataset-neutral MAS diagnosis-and-edit policy. Learn and emit the complete improvement process, not only a patch. Inspect the supplied source MAS, its real execution trajectory, and bounded node/edge diagnostics. First localize the earliest evidence-supported failure on the final-output dependency path. Then model how that failure propagated and state a falsifiable counterfactual. Finally design the smallest coherent repair plan and emit exactly one explicit operation: accept or one bounded patch operation. \revise{The input never contains the gold answer, source correctness, terminal reward, or any diagnostic derived from them; infer whether a revision is warranted only from observable execution evidence.} Do not use benchmark names, hidden gold answers, or dataset-specific answer rules as diagnostic evidence. Do not merely blame a generic planner or final synthesizer. Never emit multiple operations; the bounded patch contains exactly one edit. \revise{Allowed decisions: accept, add\_agent, rewrite\_role, add\_edge, remove\_edge, adjust\_reasoning\_budget, change\_tool. The add\_agent operation creates a task-solving agent and atomically connects its output to the current aggregation agent without changing the aggregation agent.} accept must be the only operation when chosen. Use only observed evidence; never invent execution events or copy long transcripts. Output exactly one XML <refiner> document containing stage1\_localization, stage2\_causal\_model, stage3\_repair\_plan, and patch. Escape XML special characters. No JSON, markdown, <think>, or private reasoning. Stop immediately after </refiner>.

    \end{tcolorbox}

    \caption{MAS-Loop XML Refiner prompt (Part 1/2).}
    \label{fig:mas-loop-xml-refiner-prompt-part1}
\end{figure*}

\begin{figure*}[htbp]
    \centering
    \begin{tcolorbox}[
        enhanced,
        colframe=Salmon!90!Black,
        colback=Salmon!20,
        coltitle=white,
        fonttitle=\large\bfseries,
        title={MAS-Loop Prompt for Refiner (Part 2/2)},
        halign title=left,
        fontupper=\footnotesize,
        boxrule=1pt,
        arc=3mm,
        boxsep=2pt,
        left=8pt,
        right=8pt,
        top=4pt,
        bottom=4pt
    ]

    \textbf{User\_prompt:}

    Produce one grounded three-stage improvement decision as one learned reasoning trace, not three model calls or hard filters. source\_status must be exactly one of: actionable\_failure, preserve; \revise{target\_type must be agent, edge, or mas.} Use this XML element/attribute contract. Replace every ellipsis/example, fill all elements with task-specific content, and never emit an empty <evidence><item></item> body:

    <refiner>

    \hspace*{2em}<stage1\_localization source\_status="..." target\_type="..." target\_id="..." failure\_type="...">

    \hspace*{4em}<evidence>

    \hspace*{6em}<item agent\_id="..." execution\_index="..." field="..." location="..." round="...">

    \hspace*{8em}Concrete task-specific observation from this exact trace value; replace this example.

    \hspace*{6em}</item>

    \hspace*{4em}</evidence>

    \hspace*{2em}</stage1\_localization>

    \hspace*{2em}<stage2\_causal\_model>

    \hspace*{4em}<causal\_chain>

    \hspace*{6em}<step></step>

    \hspace*{4em}</causal\_chain>

    \hspace*{4em}<causal\_explanation></causal\_explanation>

    \hspace*{4em}<counterfactual></counterfactual>

    \hspace*{4em}<downstream\_impact></downstream\_impact>

    \hspace*{2em}</stage2\_causal\_model>

    \hspace*{2em}<stage3\_repair\_plan confidence="...">

    \hspace*{4em}<repair\_goal></repair\_goal>

    \hspace*{4em}<expected\_effect></expected\_effect>

    \hspace*{4em}<risk></risk>

    \hspace*{2em}</stage3\_repair\_plan>

    \hspace*{2em}<patch>

    \hspace*{4em}<operation op="..." />

    \hspace*{2em}</patch>

    </refiner>

    Each <evidence><item> must have attributes agent\_id, execution\_index, field, and location; add round when the location references a retrieval round. location must be copied exactly from one trace\_refs value in node\_diagnostic\_state. The item text must state the concrete observed signal at that location. For an edge use agent\_id='from->to' and execution\_index='-1'. For a MAS/outcome observation use agent\_id='mas' and execution\_index='-1'. Do not copy field descriptions as observations.

    Existing agent IDs: [\{known\_agent\_ids\}]. Use these exact IDs when possible.

    Patch operation XML forms:

    \hspace*{2em}<operation op="accept" />

    \hspace*{2em}<operation op="rewrite\_role" agent\_id="ID"><role>NEW\_ROLE</role></operation>

    \hspace*{2em}\revise{<operation op="adjust\_reasoning\_budget" agent\_id="ID" budget="1" />}

    \hspace*{2em}<operation op="change\_tool" agent\_id="ID" old\_tool="OPTIONAL" new\_tool="TOOL" />

    \hspace*{2em}\revise{<operation op="add\_agent" id="NEW\_ID" budget="1"><role>ROLE</role></operation>}

    \hspace*{2em}<operation op="add\_edge" from="ID" to="OTHER\_ID" type="data\_flow" />

    \hspace*{2em}<operation op="remove\_edge" from="ID" to="OTHER\_ID" type="data\_flow" />

    Emit exactly one operation. The parser retains legacy aliases only for audit/rejection; do not use them. budget is 1--4.

    Input state is JSON data embedded only as input context; your OUTPUT must be XML:

    \{

    \hspace*{2em}"task": \{task\},

    \hspace*{2em}"constraints\_and\_capabilities": \{sanitized\_constraints\},

    \hspace*{2em}"source\_mas": \{source\_mas\},

    \hspace*{2em}"node\_diagnostic\_state": \{node\_diagnostic\_state\},

    \hspace*{2em}"optional\_sampling\_focus": \{optional\_repair\_focus\}

    \}

    \end{tcolorbox}

    \caption{MAS-Loop XML Refiner prompt (Part 2/2).}
    \label{fig:mas-loop-xml-refiner-prompt-part2}
\end{figure*}

\begin{figure*}[htbp]
    \centering
    \begin{tcolorbox}[
        enhanced,
        colframe=Salmon!90!Black,
        colback=Salmon!20,
        coltitle=white,
        fonttitle=\large\bfseries,
        title={LLM-as-a-judge Prompts (VitaBench)},
        halign title=left,
        fontupper=\footnotesize,
        boxrule=1pt,
        arc=3mm,
        boxsep=2pt,
        left=8pt,
        right=8pt,
        top=4pt,
        bottom=4pt
    ]

\hspace{0em}\textbf{System Prompt:}
System Information
\texttt{\{env\_info\}}
User Complete Instruction
\texttt{\{user\_instruction\}}

\textbf{Background}
- This is a conversation scenario between a user and an assistant, where the assistant can call tools to retrieve information and complete operations. Tool return results will start with "tool"

\hspace{4em}- You need to evaluate whether the user instruction has been completed. The user's complete instruction has been broken down into several scoring points (rubrics), and you only need to judge whether each scoring point is satisfied

\hspace{4em}- Due to the large number of conversation turns, we use sliding window evaluation, where each window shows 10 conversation turns with 2 overlapping turns between windows, but rubric status will be preserved across windows

\hspace{4em}- You are evaluating window \texttt{\{window\_idx\}} (out of \texttt{\{total\_windows\}} windows total)

\hspace{4em}- \texttt{<window\_content>} contains the conversation content for the current window

\hspace{4em}- \texttt{<current\_rubrics>} contains the current status of all scoring points (true means satisfied, false means not satisfied, all scoring points have an initial status of false)

\hspace{2em}\textbf{Task}
- Update the scoring point rubric status based on the conversation content in the current window

\hspace{4em}- You can update the status from false to true, if and only if the assistant completed the goal in this window

\hspace{4em}- You can also update true back to false, if and only if the assistant overturned a previous correct conclusion in this window (but note: if the user's needs themselves changed, such as actively canceling after placing an order, the original evaluation of the ordering requirement should not be overturned)

\hspace{4em}- You can refer to the "User Complete Instruction" to understand the progress of the current conversation window and avoid unnecessary modifications

\hspace{2em}\textbf{Important Notes}

\hspace{4em}- Important: All evaluations are based on whether the assistant's responses and tool call requests complete the goals in the rubrics. User expressions in the conversation are only considered as guidance for the assistant and do not directly affect evaluation standards. Everything is based on the rubric fields!

\hspace{4em}- Important: Query tool return results are only visible to the assistant and do not represent content recommended by the assistant to users, so they do not directly affect evaluation results either. Everything must be based on the assistant's responses to users after obtaining information! Also note that the Assistant cannot fabricate Tool return results!

\hspace{4em}- Important: For order-related rubrics (involving order details that must generate orders), you must confirm whether the assistant actually completed the ordering operation. The assistant may mistakenly believe they completed the ordering operation when in fact the tool call failed; or situations where the user states they "can place the order themselves," etc., should all be considered as not meeting the requirements

\hspace{4em}- For rubrics involving order details such as product quantity or delivery time, the original rubric requirements must be strictly met (no deviation in product quantities, delivery must not be later than the expected time). User compromise behavior does not affect evaluation results (for example, when a user states "fewer items is okay", "I have no objections to the order content" or "later delivery is fine" etc.). These situations should still be considered as not meeting the requirements

\hspace{4em}- For rubrics involving text content matching of addresses or order notes, apply the functional equivalence principle: as long as the actual content can achieve the same function (such as roughly locating the delivery location or conveying the customer's main needs), it is considered to meet the requirements even if the expression is not completely consistent or lacks some details

\hspace{4em}- If the current window does not involve a certain rubric, keep its original status unchanged; if the current window involves a rubric but cannot fully determine it, key information can be recorded in the justification for future judgment

\hspace{4em}- In the justification, record key information related to the current rubric and its corresponding round [x] in an appending manner. If status modifications occur, record the reason using concise language. If the status is not modified, restate the previous reason

\hspace{2em}Format Requirements

\hspace{4em}- Your response should be a JSON object containing the following fields:

\hspace{6em}- \texttt{rubric\_idx}: Unique identifier for the rubric

\hspace{6em}- \texttt{rubric}: Restatement of the rubric

\hspace{6em}- \texttt{justification}: Explanation of status changes

\hspace{6em}- \texttt{meetExpectation}: Updated status (true or false)

\hspace{0em}\textbf{User Prompt:}
\hspace{2em}\texttt{<window\_content>}
\texttt{\{window\_content\}}
\hspace{0em}\texttt{</window\_content>}

\hspace{2em}\texttt{<current\_rubrics>}
\texttt{\{current\_rubrics\_str\}}
\hspace{0em}\texttt{</current\_rubrics>}

    \end{tcolorbox}
    \caption{LLM-as-a-judge prompts used in VitaBench.}
    \label{fig:llm_judge_vita}
\end{figure*}

\end{document}